\documentclass[aps, prd, twocolumn,superscriptaddress, nofootinbib, amsmath, amsfonts, amssymb]{revtex4-2}

\pdfoutput=1
\usepackage[utf8]{inputenc}
\usepackage[T1]{fontenc}
\usepackage{hyperref, cleveref}
\usepackage{graphicx}

\begin{document}

\title{\texorpdfstring{Muon $g-2$ and Co-annihilating Dark Matter in the Minimal Supersymmetric Standard Model}{Muon g--2 and Co-annihilating Dark Matter in the Minimal Supersymmetric Standard Model}}

\author{Peter Cox}
\email{peter.cox@unimelb.edu.au}
\affiliation{School of Physics, The University of Melbourne, Victoria 3010, Australia}

\author{Chengcheng Han}
\email{hanchch@mail.sysu.edu.cn}
\affiliation{School of Physics, Sun Yat-Sen University, Guangzhou 510275, China}

\author{Tsutomu T. Yanagida}
\email{tsutomu.tyanagida@sjtu.edu.cn}
\affiliation{Tsung-Dao Lee Institute \& School of Physics and Astronomy, Shanghai Jiao Tong University, 200240 Shanghai, China}

\begin{abstract}
We demonstrate that the recent measurement of the anomalous magnetic moment of the muon and dark matter can be simultaneously explained within the Minimal Supersymmetric Standard Model. Dark matter is a mostly-bino state, with the relic abundance obtained via co-annihilations with either the sleptons or wino. The most interesting regions of parameter space will be tested by the next generation of dark matter direct detection experiments.
\end{abstract}

\maketitle

%============================================================================

\section{Introduction}

The recent measurement of the anomalous magnetic moment of the muon~\cite{PhysRevLett.126.141801} represents an exciting hint for the existence of physics beyond the Standard Model. The combination of the new result and the Brookhaven (E821) measurement~\cite{hep-ex/0602035} is in tension with the Standard Model prediction\footnote{This uses the $R$-ratio method~\cite{1810.00007,1907.01556,1908.00921,1911.00367} for the HVP contribution. A recent lattice determination~\cite{Borsanyi:2020mff} yields a different value which, if correct, would alleviate the tension in \cref{eq:amu}.}~\cite{2006.04822} at $4.2\sigma$:
\begin{equation} \label{eq:amu}
    a_\mu^\text{exp} - a_\mu^\text{SM} = (2.51 \pm 0.41_\text{(exp)} \pm 0.43_\text{(theory)} ) \times 10^{-9} \,.
\end{equation}

Supersymmetry is one of the leading candidates to explain this discrepancy. In this letter, and in light of the new measurement, we explore the possibility that the Minimal Supersymmetric Standard Model (MSSM) could be responsible for both the deviation in the muon $g-2$ and the dark matter (DM) of the universe.

The MSSM features a natural dark matter candidate in the form of the lightest neutralino. It is by now well understood that the most promising scenario to simultaneously explain the deviation in $a_\mu$ and obtain the observed relic abundance via thermal production involves bino-like dark matter and co-annihilations~\cite{1704.05287,1710.11091,1805.02802,1909.07792,2006.15157}. This conclusion is primarily driven by the strong bounds from dark matter direct detection, combined with the fact that it is not possible to obtain a large enough contribution to $a_\mu$ for either pure Higgsino or Wino dark matter. 

There are two distinct scenarios to consider, depending on the identity of the co-annihilating partner: \emph{bino-slepton} and \emph{bino-wino} co-annihilation. As we shall demonstrate, both of these scenarios have regions of parameter space that can explain the result in \cref{eq:amu} while simultaneously accounting for the dark matter relic abundance and evading all other constraints.

%============================================================================

\section{Analysis Framework and Assumptions}
\label{sec:assumptions}

We begin by describing the details of our analysis. First, we assume that the squarks, gluinos and additional Higgs bosons are all decoupled, motivated by the strong bounds from collider searches. For concreteness we fix their masses to be $\sim3$\,TeV. This is also the renormalisation scale at which all the parameters are specified. The trilinear coupling $A_t$ is fixed to 5\,TeV in order to obtain a Higgs mass of $\approx125$\,GeV, and we assume that all other $A$-terms are negligible. 

To simplify our analysis, we assume that the left and right-handed slepton soft masses are equal. We also take equal slepton soft masses for the first and second generations. We focus on the case where $\text{sgn}(M_{1,2}\mu)>0$, since this ensures that the dominant contributions to $a_\mu$ have the correct sign to account for the difference in \cref{eq:amu}.

We use the spectrum generator {\tt SuSpect-2.52}~\cite{hep-ph/0211331}, while the dark matter relic abundance, DM-nucleon scattering cross-section, and one-loop SUSY contributions to $a_\mu$ are all calculated using {\tt MicrOMEGAs-5.2.7}~\cite{2003.08621}. We also include the leading two-loop contributions to $a_\mu$. These come from  $\tan\beta$-enhanced corrections to the muon Yukawa coupling~\cite{hep-ph/9912516,0808.1530} and the QED running down to the muon mass scale~\cite{hep-ph/9803384}.

%============================================================================

\section{Bino--stau co-annihilation}
\label{sec:bino-stau}

\begin{figure*}[t]
    \centering
    \includegraphics[width=0.49\textwidth]{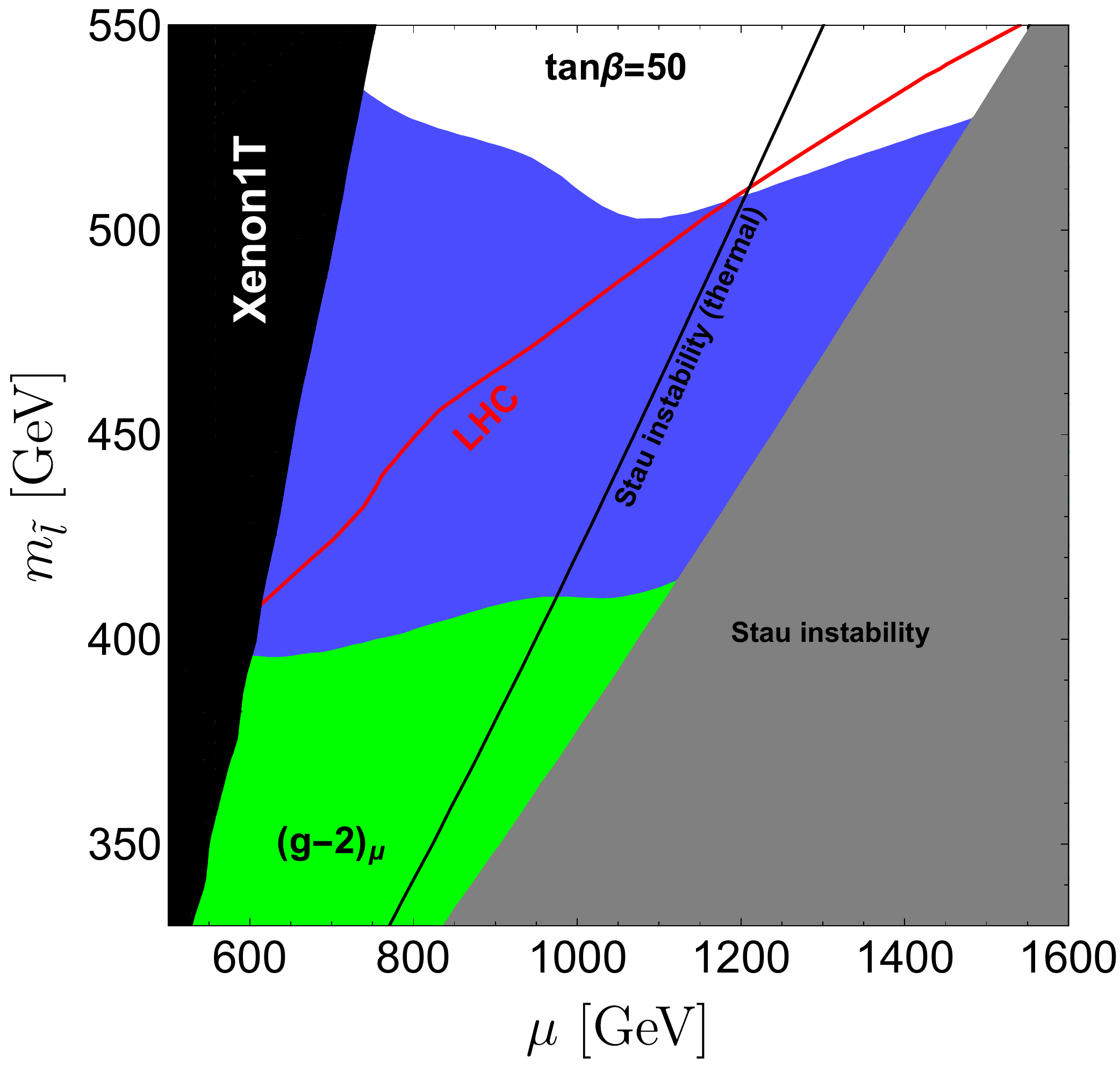}
    \includegraphics[width=0.49\textwidth]{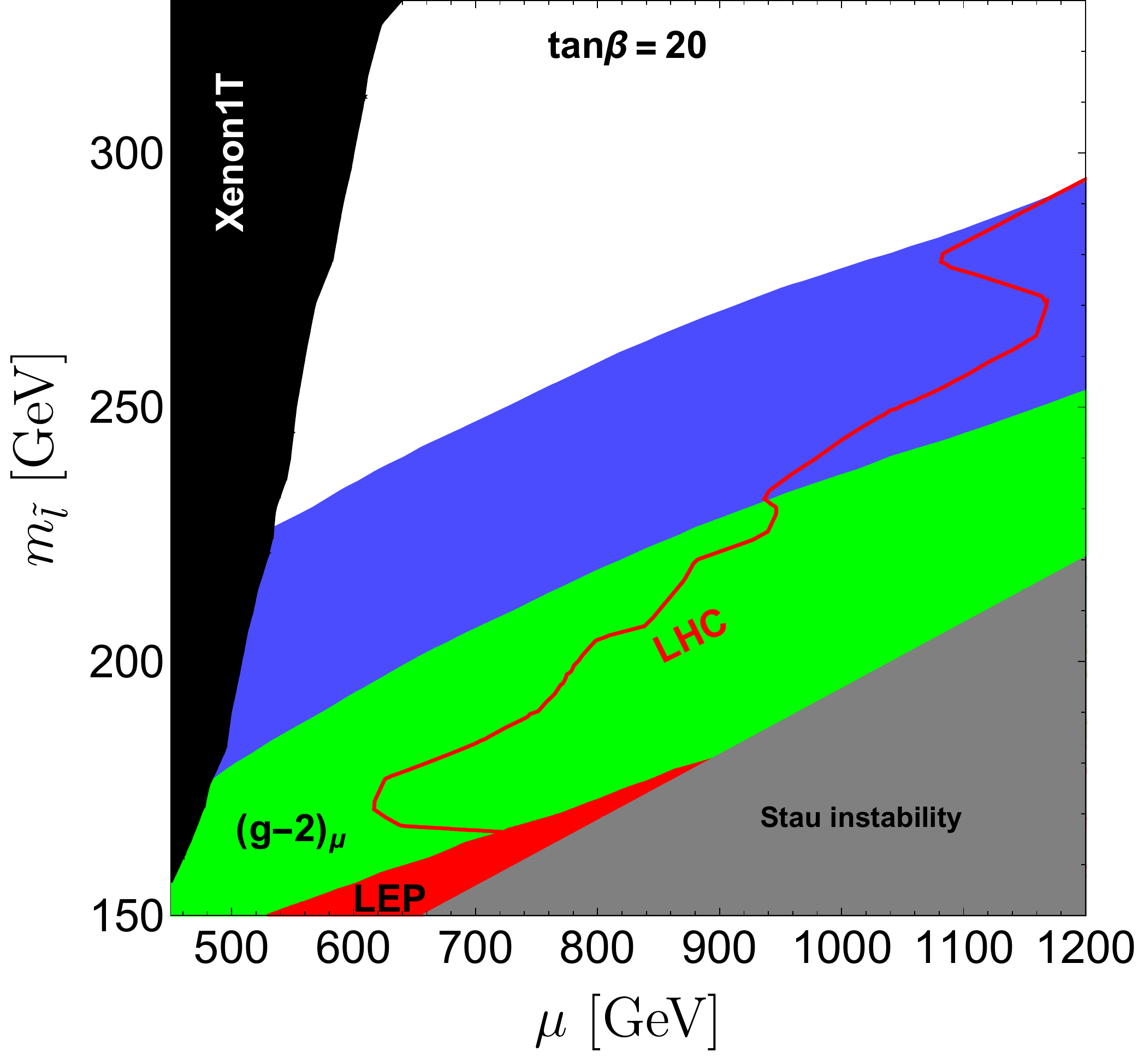}
    \caption{Bino--stau co-annihilation. $M_1$ has been adjusted to obtain the correct relic abundance, and we fix $M_2=1$\,TeV. The green (blue) region is consistent with $a_\mu$ at $1\sigma$\,($2\sigma$). The black region is excluded by XENON1T~\cite{1805.12562}. The region to the right of the red line is excluded by the ATLAS slepton search with $139\,\text{fb}^{-1}$~\cite{1908.08215}, and the red region is excluded by slepton searches at LEP~\cite{LEP-slepton}. The grey region is excluded by vacuum instability. Left: $\tan\beta=50$. Right: $\tan\beta=20$.}
    \label{fig:bino-stau}
\end{figure*}

The first scenario we consider is bino-slepton co-annihilation with universal slepton masses. The assumption of universal slepton masses is often imposed in order to avoid dangerous contributions to flavour changing neutral current (FCNC) processes. It can be motivated by certain supersymmetry breaking scenarios, such as gaugino mediation~\cite{Inoue:1991rk,hep-ph/9911293,hep-ph/9911323}. In this case, the NLSP and co-annihilating partner is the lightest stau. Achieving the correct relic abundance then requires a mass-splitting  $m_{\tilde{\tau}_1} - m_{\chi^0_1} \lesssim15$\,GeV.

The stau co-annihilation region is shown in \cref{fig:bino-stau}, where we have taken $M_2=1$\,TeV. $M_1$ is adjusted across the parameter space in order to obtain the correct relic abundance. Consider first the left panel where we fix $\tan\beta=50$. The green (blue) regions fit the $a_\mu$ measurement at $1\sigma$ ($2\sigma$).  However, the $1\sigma$ region is already excluded by LHC searches for the first and second generation sleptons~\cite{1908.08215,CMS-PAS-SUS-19-012} (red line). 

There is also a constraint from vacuum stability, due to the existence of charge-breaking minima in the scalar potential when $\mu\tan\beta$ becomes large~\cite{1011.0260}. We take the bound from Ref.~\cite{1809.10061}; the grey region is excluded at zero temperature, while a stronger bound (black line) is obtained by considering the finite temperature effective potential and requiring stability throughout the thermal history of the universe (note that this assumes a sufficiently high reheating temperature).

The right panel of \cref{fig:bino-stau} corresponds to $\tan\beta=20$.  The smaller value of $\tan\beta$ has the effect of compressing the slepton spectrum, which relaxes the bounds from LHC slepton searches. There are then regions which can fit $a_\mu$ at $1\sigma$, while evading the bounds from collider searches. These correspond to very small slepton masses, with the lightest stau close to the LEP lower bound. 

It is clear from \cref{fig:bino-stau} that bino-stau co-annihilation with universal slepton masses, while still viable, is strongly constrained. One could consider decreasing $M_2$ in order to increase the chargino-sneutrino contribution to $a_\mu$. This would move the best-fit region towards larger slepton masses and away from the collider bounds. However, $M_2$ cannot be decreased significantly without encountering bounds from chargino searches, particularly from slepton-mediated decays~\cite{1908.08215}. 

While LHC searches currently provide the strongest constraints on this scenario, dark matter direct detection will have an important role in the future. The entirety of the best-fit regions in both panels of \cref{fig:bino-stau} will be probed by the LZ experiment~\cite{1802.06039}. Indirect detection is not currently sensitive to any of the co-annihilation scenarios, due to the small annihilation cross-section today. In the future, the Cherenkov Telescope Array should have sensitivity to the bino-stau co-annihilation scenario~\cite{2007.16129}.

Finally, note that we are assuming universal slepton soft masses at low-scale. In a UV model this relation might be expected to hold at high scales, but the RG running tends to reduce the stau soft mass compared to the first and second generation sleptons. This may lead to slightly stronger bounds from slepton searches for small $\tan\beta$, but we do not expect this effect to significantly alter our conclusions (see also~\cite{1811.12699}).

%============================================================================

\section{Bino--slepton co-annihilation}
\label{sec:bino-slepton}

\begin{figure*}[t]
    \centering
    \includegraphics[width=0.49\textwidth]{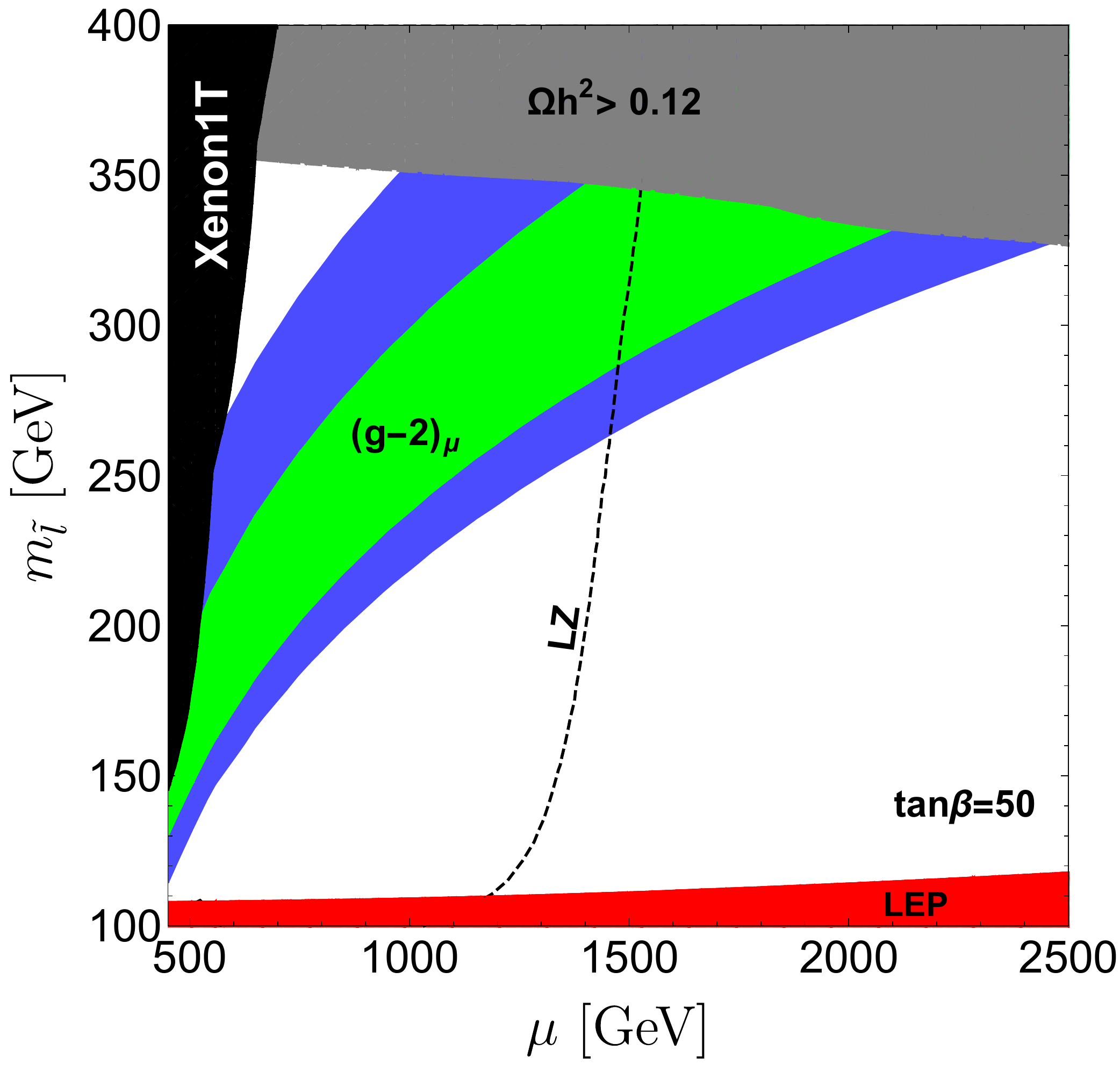}
    \includegraphics[width=0.49\textwidth]{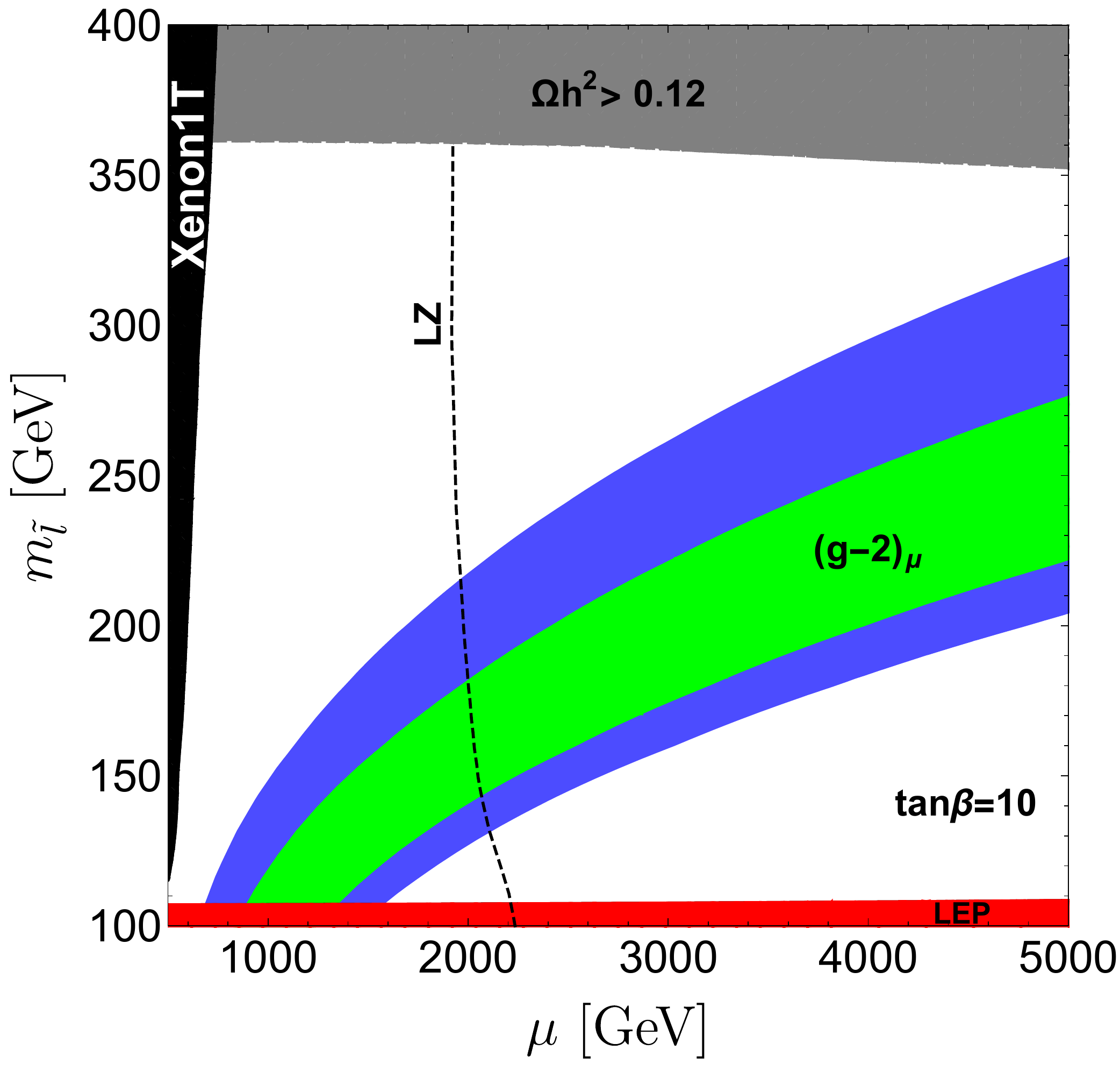}
    \caption{Bino--slepton co-annihilation. We have fixed $\tan\beta=50$ (left) and $\tan\beta=10$ (right). $M_1$ has been adjusted to obtain the correct relic abundance; for large slepton masses this becomes impossible as shown by the grey region. The green (blue) region is consistent with $a_\mu$ at $1\sigma$\,($2\sigma$). The black region is excluded by XENON1T~\cite{1805.12562}, while the dashed line shows the future sensitivity of the LZ experiment~\cite{1802.06039}. The red shaded region is excluded by slepton searches at LEP~\cite{LEP-slepton}. }
    \label{fig:bino-slepton}
\end{figure*}

Given the results of the previous section, we now relax the assumption of flavour universality for the slepton masses and assume that the staus are decoupled ($m_{\tilde{\tau}_L}=m_{\tilde{\tau}_R}=3$\,TeV). As we shall see, this opens up significant regions of parameter space in which both the muon $g-2$ and dark matter can be accommodated. This type of spectrum can be realised, for example, in Gaugino+Higgs mediation~\cite{1811.12699}\footnote{This is achieved with large negative soft masses for the Higgs doublets at the GUT scale, and $\mu\sim\mathcal{O}(10)$\,TeV~\cite{1606.04953,1607.05705}.}, without inducing large FCNCs~\cite{1801.05785}. 

In this scenario the NLSP is the lightest smuon. The results are shown in \cref{fig:bino-slepton} for $\tan\beta=50\,(10)$ in the left (right) panels. $M_1$ has been adjusted to obtain the observed relic abundance, which requires a mass-splitting $m_{\tilde{\mu}_1}-m_{\tilde{\chi}^0_1} \lesssim 20\,$GeV. The required mass-splitting decreases for larger slepton masses, and there is an upper bound on the slepton mass (grey region) above which it is no longer possible to obtain the correct relic density. When $\mu \tan\beta$ is small, the lightest selectron and smuon are almost degenerate and both act as co-annihilation partners. As $\mu\tan\beta$ is increased, $\tilde{\mu_1}$ becomes lighter due to the left-right mixing and plays the dominant role in setting the relic abundance. This also leads to a decrease in the upper bound on the slepton mass.

Here, we have assumed that the wino is decoupled ($M_2=3\,$TeV); the dominant contribution to $a_\mu$ therefore comes from the bino-smuon loop. This contribution is proportional to the left-right smuon mixing and is enhanced for large $\mu\tan\beta$. Hence, the best-fit region for $a_\mu$ moves towards higher slepton masses as $\mu$ is increased. For large $\tan\beta$, much of the best-fit region will be probed in the relatively near future by dark matter direct detection. Reducing $\tan\beta$ has the effect of moving the best-fit region for $a_\mu$ to lower slepton masses for a given $\mu$. This is seen in the right panel of \cref{fig:bino-slepton}, where there is significant viable parameter space well beyond the reach of future direct detection experiments. 

We now briefly discuss collider searches for the light sleptons. The compressed spectrum makes this scenario challenging to probe at hadron colliders. Nevertheless, there is a dedicated ATLAS search targeting bino-slepton co-annihilation~\cite{1911.12606}. However, it does not currently constrain the parameter space in \cref{fig:bino-slepton}. Given the upper bound on the slepton mass from the relic abundance, the bino-slepton co-annihilation scenario could, however, be fully tested at a future lepton collider with $\sqrt{s}>700\,$GeV (see Refs.~\cite{1812.02093,2004.02834}).

%============================================================================

\section{Bino--wino co-annihilation}
\label{sec:bino-wino}

\begin{figure}[t]
    \centering
    \includegraphics[width=0.49\textwidth]{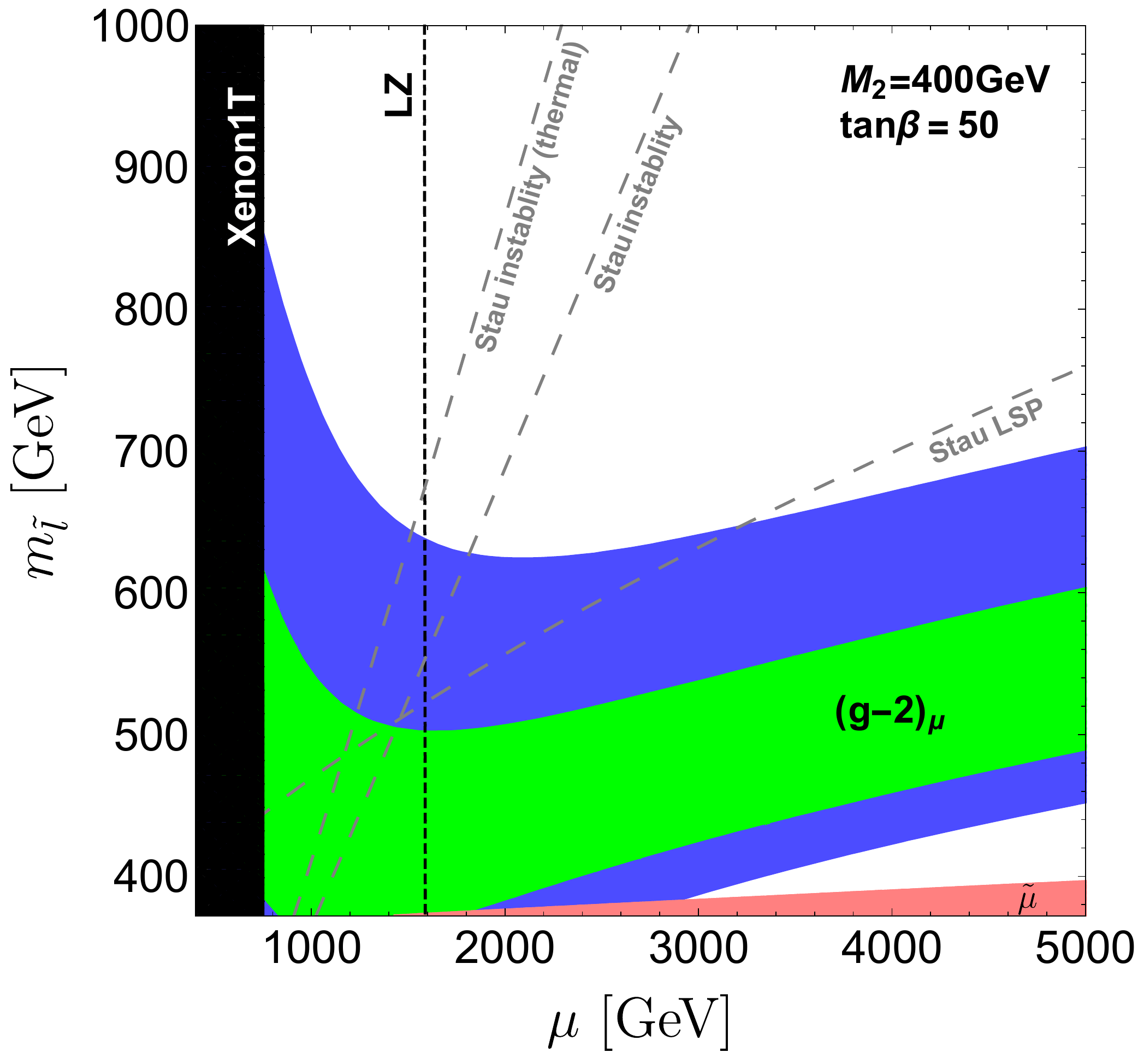}
    \caption{Bino--wino co-annihilation. We have fixed $\tan\beta=50$ and $M_2=400$\,GeV, with $M_2-M_1=28$\,GeV to obtain the correct relic abundance. The LSP mass is $m_{\tilde{\chi}^0_1}\approx370$\,GeV. The green (blue) region is consistent with $a_\mu$ at $1\sigma$\,($2\sigma$). The black region is excluded by XENON1T~\cite{1805.12562}, while the black dashed line shows the future sensitivity of the LZ experiment~\cite{1802.06039}. In the pink region $\tilde{\mu}_1$ becomes the LSP. The grey dashed lines apply only for universal slepton masses, in which case the region to the right is excluded.}
    \label{fig:bino-wino}
\end{figure}

Last, we consider bino-wino co-annihilation. In this scenario a slightly larger mass-splitting of $10-30$\,GeV is needed between the bino-like LSP and wino-like NLSP~\cite{1311.2162,1403.0715}. This situation is shown in \cref{fig:bino-wino} for $M_2=400$\,GeV and $\tan\beta=50$. We have fixed $M_2-M_1=28$\,GeV, which gives approximately the correct relic abundance across the parameter space (the precise mass-splitting needed has a mild $\mu$-dependence). There are significant regions of parameter space that can explain both $a_\mu$ and the dark matter abundance. For large $\mu$, the bino-smuon contribution to $a_\mu$ dominates, while for $\mu \lesssim1.5$\,TeV the chargino-sneutrino contribution becomes important and moves the $g-2$ best-fit region to higher slepton masses. This latter region will be probed by future direct detection experiments

Collider searches for the sleptons and wino can potentially provide powerful probes of this scenario, although do not currently constrain the parameter space shown in \cref{fig:bino-wino}. Both ATLAS~\cite{1911.12606} and CMS~\cite{CMS-PAS-SUS-18-004} have dedicated chargino searches targeting the compressed spectra relevant for bino-wino co-annihilation. Currently, the CMS search with $137\,\text{fb}^{-1}$ obtains a bound of $m_{\tilde{\chi}^0_2/\tilde{\chi}^1_\pm} > 280$\,GeV for $\Delta m = 10\,$GeV, reducing to $m_{\tilde{\chi}^0_2/\tilde{\chi}^1_\pm} > 200$\,GeV for $\Delta m = 30\,$GeV. These limits assume that $\tilde{\chi}^0_2$, $\tilde{\chi}^1_\pm$ decay purely via off-shell gauge bosons. Here, there can also be decays to leptons mediated via the light sleptons, which strengthens the bounds slightly. At the HL-LHC these searches are projected to eventually be sensitive to $m_{\tilde{\chi}^0_2/\tilde{\chi}^1_\pm}\lesssim 430$\,GeV~\cite{1804.05238}. 

Slepton searches can already probe masses up to 700\,GeV~\cite{1908.08215,2012.08600}, although the limits become weaker with increasing LSP mass and there is currently no bound for $m_{\chi^0_1}>400$\,GeV. For smaller values of $M_2$ than shown in \cref{fig:bino-wino}, slepton searches do constrain the parameter space. However, care should be taken when imposing the limits. First, the naive limit is weakened by the fact that the $\text{BR}(\tilde{\ell}^\pm\to\chi^0_{1,2} \ell^\pm)\approx60\%$, since the decay mode $\tilde{\ell}^\pm\to\chi^\pm_1\nu$ is also accessible. Second, in addition to slepton pair production, the processes $pp \to \tilde{\nu} \tilde{\ell}$ and $pp \to \tilde{\nu} \tilde{\nu}$, with $\tilde{\nu} \to \chi^\pm_1 \ell^\mp$, may also pass the analysis cuts. This is because the additional soft leptons or jets from the subsequent chargino decay may not be reconstructed. This has the potential to strengthen the limit due to the large $\tilde{\nu} \tilde{\ell}$ production cross-section, but requires a full recasting of the analysis. 

In the future, slepton searches at the (HL-)LHC are expected to probe most of the $2\sigma$ region for $a_\mu$, with two exceptions. The first is the fully compressed region where $M_1 \approx M_2 \approx m_{\tilde{\ell}}$. The second is when $\mu$ becomes extremely large, in which case $a_\mu$ can be explained with smuon masses exceeding $1$\,TeV~\cite{1309.3065}.

So far in this section we have assumed that the staus are decoupled. Let us now briefly comment on the case of universal slepton masses. In this case the large-$\mu$ region is excluded due to the bound from vacuum stability or because the lightest stau becomes the LSP (grey dashed lines). The small surviving region that can explain $a_\mu$ will soon be tested by direct detection, although for smaller values of $\tan\beta$ this region moves beyond the reach of LZ. 

%============================================================================

\section{Conclusion}

The discrepancy between the measurement of the anomalous magnetic moment of the muon and its Standard Model prediction now exceeds $4\sigma$, providing a tantalising hint for physics beyond the Standard Model. In this letter, we have demonstrated that this result can easily be accommodated within the framework of the MSSM, with a bino-like LSP simultaneously responsible for dark matter. The observed relic abundance is achieved through co-annihilations with either the sleptons or a light wino. 

We find that bino-stau co-annihilation with universal slepton masses is now strongly constrained by LHC searches and will be thoroughly tested by the LZ experiment. On the other hand, with non-universal slepton masses the majority of the best-fit region for $a_\mu$ currently remains unconstrained for both the bino-slepton and bino-wino co-annihilation scenarios. 

In both scenarios, the regions with $\mu\lesssim1.5$\,TeV will be probed by the next generation of dark matter direct detection experiments in the near future. This is especially interesting, given that this region is also theoretically preferred by naturalness.

%============================================================================

\section{Acknowledgements}

The authors would like to thank Norimi Yokozaki for useful discussions. The work of P.C. was supported by the Australian Government through the Australian Research Council. C.H. is supported by the Guangzhou Basic and Applied Basic Research Foundation under Grant No. 202102020885, and the Sun Yat-Sen University Science Foundation. T.T.Y. is supported by China Grant for Talent Scientific Start-Up Project.

%=============================================================================
\bibliography{main}

\end{document}